# Cosmic Muon Detector Using Proportional Chambers

**Dezső Varga[a], Zoltán Gál[b], Gergő Hamar[a], Janka Sára Molnár[c], Éva Oláh[b,d], Péter Pázmándi[b]**

[a] MTA Wigner Research Centre for Physics, Budapest; [b] Mechatronika Secondary School, Budapest; [c] Teleki Blanka Secondary School, Székesfehérvár; [d] Eötvös Loránd University, Budapest

Email: Varga.Dezso@wigner.mta.hu

**Abstract**

A set of classical multi-wire proportional chambers were designed and constructed with the main purpose of efficient cosmic muon detection. These detectors are relatively simple to construct, and at the same time are low cost, making them ideal for educational purposes. The detector layers have efficiencies above 99% for minimum ionizing cosmic muons, and their position resolution is about 1 cm, that is, particle trajectories are clearly observable. Visualization of straight tracks is possible using an LED array, with the discriminated and latched signal driving the display. Due to the exceptional operating stability of the chambers, the design can also be used for cosmic muon telescopes.

## 1. Introduction

The invention of the Multi-Wire Proportional Chamber (MWPC) by G. Charpak [1] in the 1960s heralded the era of "electronic detectors" in high energy particle physics, an achievement for which he was awarded the Nobel Prize for Physics in 1992. The construction and operation of such detectors and their derivatives (drift chambers and time projection chambers) are discussed extensively in textbooks such as that of Rolandi and Blum [2], and the CERN lecture notes by Sauli [3]. Five decades of experience has helped designers successfully construct MWPCs and avoid simple or tricky traps. Such detectors were used from the revolutionary years of 1970s, and are still considered as baseline solutions, only to be superseded in most parameters in the last decade by micro-pattern gaseous detectors [4].

An MWPC is based on a set of parallel thin anode wires. Close to the wires on positive potential, the field strength increases so as to initiate a Townsend-avalanche: electrons become energetic enough to ionize the gas, and in turn grow exponentially in numbers. The gas filling is usually 60-90% noble gas, most notably argon, whereas additional molecular gas needs to be mixed, such as $CO_2$ or methane, to ensure stable avalanche formation. The gas needs to be free of oxygen: in fact 0.1% of $O_2$ in the gas reduces the sensitivity drastically. MWPCs are sensitive to all forms of ionizing radiation. High energy charged particles ionize the gas with typically 100 electrons liberated along the path of 1cm length, which are in turn collected and amplified on the anode wires.

The wires in an MWPC are special in the sense that they must be thin enough to produce the sufficiently high electric field close to their surface, and at the same time need to be strong enough to ensure mechanical stability. Experience shows that gold-plated tungsten wires of 15-40 μm diameter are reliably usable, with the optimal range being 20-30 μm for high gain, argon-based atmospheric detectors. Other wires may serve as field shaping electrodes to optimize the field line distribution inside the chamber, and such wires should be thicker with 100-200 μm copper or brass wires being optimal.

MWPCs are usually constructed from relatively inexpensive materials. As the inner structure is complicated to ensure wire geometry, the materials are rarely good in terms of maintaining gas quality. For this reason, the gas inside the detector is continually, slowly flushed, with typical 2-10 hours of complete volume change. This means that gas supply needs to be maintained with a constant, low flow, in order of 0.5 – 5 liters per hour. With a single standard high-pressure gas bottle, however, detector systems can operate continuously for many months. Typical construction materials are glass-reinforced epoxy (same as that used for printed circuit boards, with or without copper layer), aluminium, polyethylene, Mylar or Plexiglas (PMMA).

The detector chamber is most often glued using a two-component epoxy resin, for its strength, chemical stability and constant volume during the curing process.

In the present paper, we describe the construction of a set of MWPCs, based on the combination of classical experiences extensively documented in earlier publications, and the possibilities offered by presently accessible technology. The design can be, with competent supervision, adapted for less experienced groups or even undergraduate / secondary school students. Some of these detectors were actually built by students in collaboration with scientists. The paper, being not able to accommodate all details, attempts to give an overview – the Reader is kindly asked to refer to the above mentioned textbook literatures [2,3] for detailed explanation of these standard procedures or methods.

**2. Chamber construction**

The detector design is based on the classical experiences in high energy physics, as well as our own development work [6]. The key is the simplicity of the design, which ensures reliable detector performance even if built by an inexperienced group. The material choice is also following this line, being cost efficient and ensuring high success rate in detector building. Unlike many classical MWPC-s, the detector is a firmly glued box without the possibility of repair in case of construction error or damage – according to experience the gain in the simplicity of building balances the loss of broken units. Furthermore, the details of the procedures described below may be refined during the actual work, for this reason only the most relevant ideas and possible critical issues are addressed.

The wire geometry was chosen such that anode wires are separated by 12mm, and in between field shaping wires were placed. The gas gap was 20mm. The anode wires, on positive high voltage, are 24 µm thick, whereas the field shaping wires, on ground potential, are 100 µm. The cross section of the detector is shown in Figure 1, showing the anode (sense) wires as well as the field shaping wires, with the two grounded cathode planes defining the sensitive volume. The field lines, shown in the right panel of Figure 1 using the Garfield simulation [5], are pointing towards the anode wires, hence guiding the electrons towards the amplification region. The field shaping wires actually do not strongly modify the field structure relative to the case without such wires (note there are few field lines emerging), but in this design the field wires also act as signal pick-up sensors for position information, as discussed in Section 5.

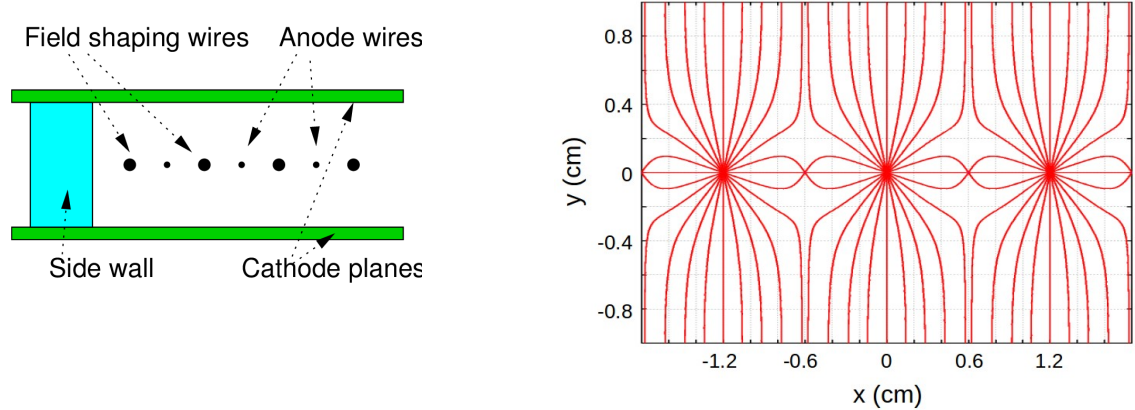

Figure 1. Left: Cross section of the chamber, close to the side wall, with wires running perpendicular to the plane of the image. Right: Electric field lines inside the chamber, which guide the electrons towards the anode wires (at x=-1.2, 0 and +1.2) where the avalanche amplification takes place.

The wire positioning and tensioning, as well as electric connections, are usually the most complicated parts of an MWPC. For the presented design the side view of the wire fixation is shown in Figure 2. The wires are positioned using grooves in appropriately shaped Plexiglas (alternatively glass-reinforced epoxy) bars on both ends, onto which a supporting printed circuit board (PCB) is glued. Once the wires are stretched over the detector box, the fixing is done by soldering on the specified spots of the PCB. Wire positioning needs to be precise to at least 0.3mm, this is the reason for using the fine grooves in the Plexiglas support bar; alternatively if the soldering is precise enough one can design a simpler structure.

After preparing the components and gluing the wire fixing bar to one of the glass-reinforced epoxy cathode plane (same material as common PCB-s), the key (and for students, the most inspiring) construction step is the wire stretching. For particle physics detectors this action has a broad literature, and needs to be done with considerable care. The wire tension for a typical 25 micron thick tungsten wire can be 15-30g (0.15 – 0.3N), and for the 100 micron copper or brass field shaping wires 50 – 100g is optimal. With the presented design, no drawbacks of uneven wire tension was experienced (neither loose wires sagging nor too tight wires breaking).

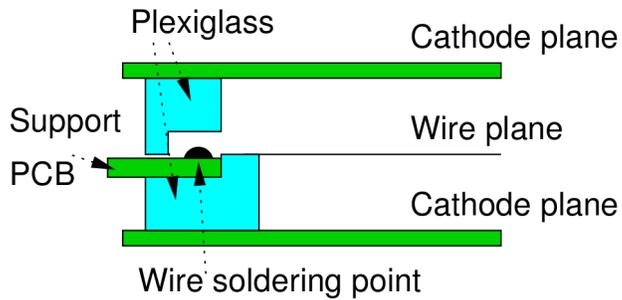 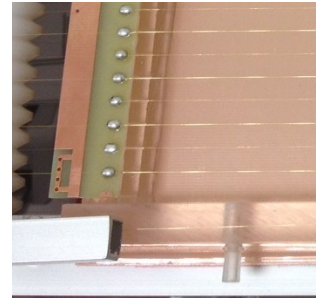

Figure 2. Left: Cross section of the detector chamber, showing the wire fixing by soldering and the surrounding structure to ensure gas tightness. All these parts are glued. Right: close-up view of the wire fixing (just before cutting the wires)

Wire stretching was done by a "winding" procedure, where a tool in a shape of a frame was used to fix two chambers at a time and bring the wires around. The system is illustrated in Figure 3 showing the path of the wire as well as a side view of the rotating frame. Figure 3 shows an image of the stretching tool from an other aspect. First always the less vulnerable field shaping wires were stretched, followed by the anode wires. The wire was tensioned by an electric motor, tuned to produce an approximately constant torque, and having the wire spool fixed on the motor axis. Wire tension calibration was done by hanging a fixed weight on the wire and setting the motor current accordingly.

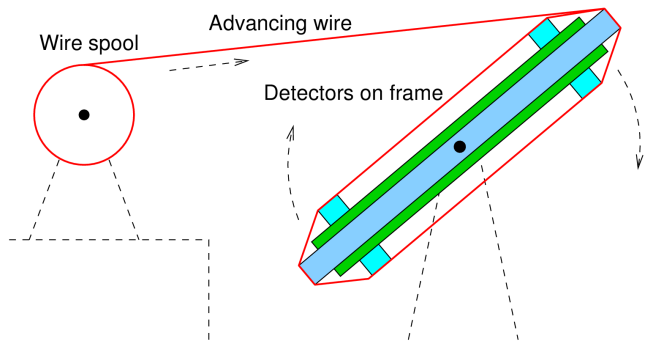 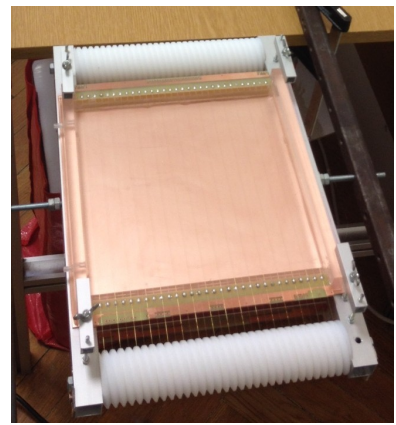

Figure 3. Left: Wire stretching tool seen from the side, showing the rotating frame with two attached detectors. Right: image of the wire stretching, with one of the chambers attached visible. The mounted structure is rotated around the axis in the middle.

Once the wires are stretched on the chambers, each of the fixing points needs to be soldered to permanently mount the wires, as discussed above. After soldering, the wire ends is to be very carefully cut: even a small wire end sticking out of the soldering spot can induce corona discharges, and thus compromising detector performance.

The last step of chamber construction is to close the gas volume. The 2cm high side walls are to be glued on the cathode plane (either before or after the wire stretching), and the chamber closed with the top cathode plane. Small leaks main remain if building is done by inexperienced people, which can be repaired by filling with glue. The gas in- and outlets can be conveniently fixed to holes on the side walls.

Once the leak tightness of the detector chamber is confirmed, the electronic connections need to be prepared. The basic circuit diagram of the high voltage feed and of the signal extraction is shown in Figure 4. All this can be conveniently installed on the properly designed wire fixing PCB, as shown in Figure 2 and Figure 5. The circuitry which feeds high voltage to the anode wires must filter noise from the input HV cable, whereas the anode signal needs to be coupled through a HV capacitor towards the pre-amplifier. The value of the of the signal coupling capacitor must be much larger than the input capacitance of the connected amplifier: as this latter is typically 10-100pF, a HV rated capacitor of a few nF is to be chosen. The noise filter resistor must be such that the mean voltage drop should be below an order of 1V: even at very high anode currents such as 100nA, this allows resistors as large as 10MΩ to be used. It is useful to choose the noise filtering cut-off frequency, given by $1/(2\pi RC)$, to be below 50-60Hz to reduce pick-up from the power lines, therefore few nA in combination with the 10MΩ is reasonable. With this setup, the detector can be tensioned with standard laboratory high voltage power supplies, which can supply 10 µA up to 2kV.

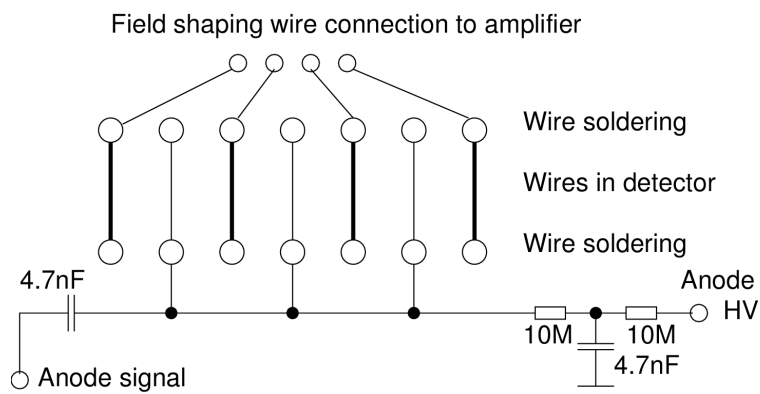

Figure 4. Electronic connection of the anode high voltage, as well as coupling of the anode signal. The T-filter on the right, formed by two 10MΩ resistors and a capacitor suppresses the high frequency noise, whereas the (fast) anode signal can be picked up, relative to ground, on the left. Both capacitors must be HV rated.

In the presented design, the anode wires are on positive high voltage, and all other electrodes are on ground potential. Good grounding is actually very important to achieve a good noise performance: the cathode planes need to be connected with each other with short, soldered, low resistance wires at least at all four edges of the chamber, and also to the grounding on the wire fixing PCB-s.

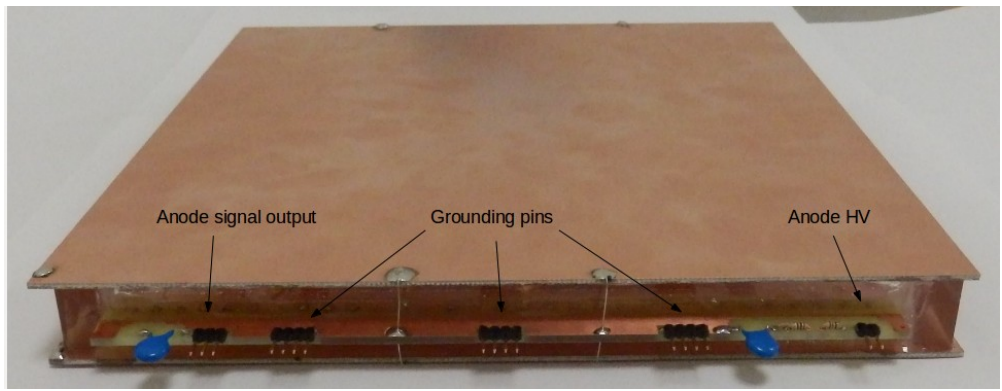

Figure 5. Completed MWPC detector. The anode HV feed (right side) and the anode signal coupling (left side) are well visible, so as the grounding wires connecting the wire fixing PCB and the top / bottom cathode planes.

## 3. Good practices for first live tests and operation

The detector may be operated with most gases used for MWPCs, however the optimal solution may be the mixture of Argon and $CO_2$, which is a cost-efficient, non-flammable, non-toxic mixture. Mixture ratios can be between 70:30 to 90:10; we report here for Ar:$CO_2$ mixture ratios of 82:18. Actually such gas is used in the welding industry as shielding gas, therefore may be easily accessible. The gas flow rate needs to be sufficient to maintain reasonable gas purity: with leak-free detectors a gas flow of below 0.5 litres / hour is sufficient.

During the first test of a constructed detector, the gas flow needs to be maintained for some time to achieve about 10 times the volume change inside the detector in order to reduce the level of residual Oxygen sufficiently. Bringing up the high voltage is usually a fast process, however one has to watch the current drawn on the HV line: currents above 100nA signals a malfunctioning detector. Typical values of current on the anode HV were below 10nA after 3 hours of operation, after the first time of switching on the detectors. With this specific setup, a total gain of $10^4$ is reached at 1600V.

At about 80% of the final operating voltage, signals from a radioactive source, or from cosmic rays, should appear. The signal may be picked up on the anode wire coupling capacitor. The pre-amplifier connected here (possibly combination of charge sensitive pre-amplifier and a shaper) optimally has about 1 – 2 µs pulse width and input-equivalent noise below 10000e (1fC): see typical oscilloscope shots in Figure 6. Excessive noise is usually a sign of improper grounding or electrical shielding.

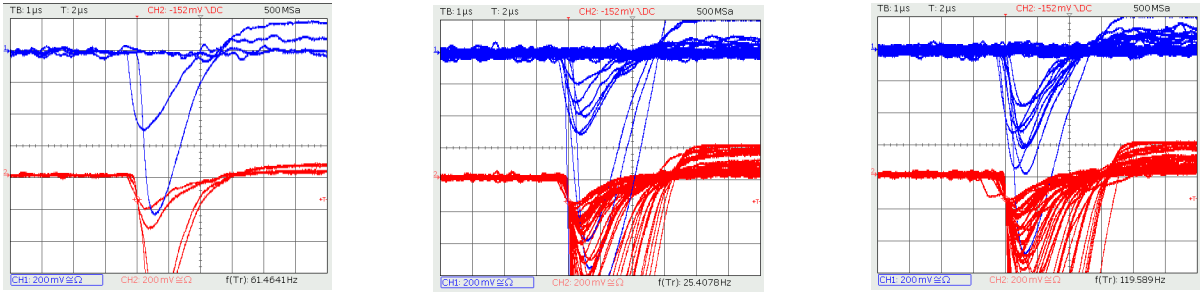

Figure 6.: Typical MWPC anode signals viewed by an oscilloscope, using a pre-amplifier with about 1 µs peaking time. The signals from one detector (red, lower trace) are used as a trigger, whereas the appearing coincident signals from the other one (blue, upper trace) are resulting from cosmic particles. Note clear separation of signal and noise, in this case at 1600V.

## 4. Detector performance studies using radioactive sources

Detector performance has been verified using radioactive sources. $^{90}$Sr emits β rays (1 – 2 MeV electrons) whereas $^{55}$Fe is an X-ray (electromagnetic photon of 5.9 keV energy) source. The responses for these sources are rather different. β rays can cross the detectors and be tagged, leaving only a few keV energy deposited inside the sensitive volume. The 5.9 keV photons on the other hand deposit their full energy by photo-effect, resulting in a sharp, well defined signal amplitude.

The measurement setup for β rays is shown in Figure 7. The electrons crossing the detector are also detected in a scintillator. In case of signals observed in the scintillator, the pulse height from the MWPC is recorded by an ADC (analogue-digital converter). The result is shown on the right panel of Figure 7: a clear separation of a sharp "noise" peak (electronic noise or cosmic rays which miss the MWPC) and signal from traversing β particles with a broad structure. The measurement was repeated at different anode wire voltages, which demonstrated that the signal amplitude scales with the avalanche gain.

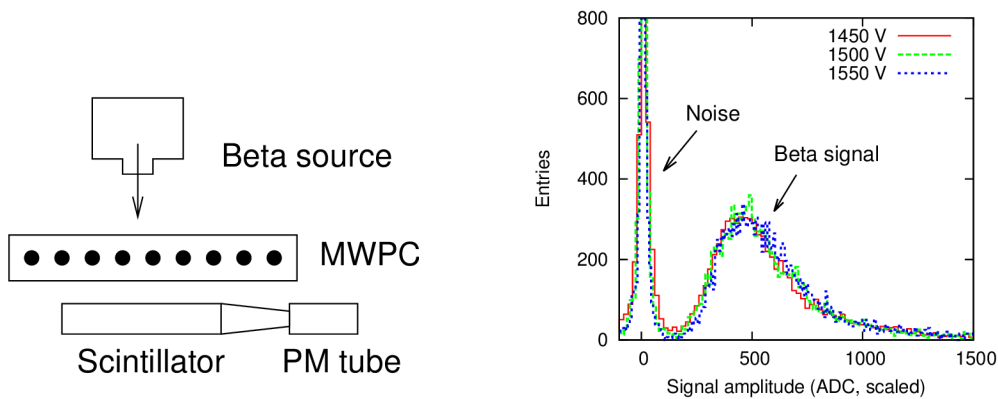

Figure 7. Measurement setup using a β source, $^{90}$Sr, in which the electrons cross the detector and are also tagged in the scintillator. The amplitude of the signal in the MWPC is shown on the right, measured in case of a signal in the scintillator, with "noise" and "particle signal" well separated. Data at different anode voltages are scaled horizontally to demonstrate shape similarity.

The response for low energy X-rays, as expected, results in a sharp (Gaussian) peak at the photon energy. Using an $^{55}$Fe source, which emits 5.9keV photons (and also creates some fraction of 3keV deposits in argon), the pulse height distribution is shown in Figure 8. This feature allows one to reliably determine the gas amplification G, which is defined as the average number of electrons in an avalanche initiated by a single electron. Measuring the X-ray detection frequency *f* (count rate) as well as the anode current *I*, and knowing that each 5.9keV photon deposit approximately 220 electrons in argon-rich Ar-$CO_2$ mixture [4], one can determine the gas gain *G*:

$$G = \frac{I}{f * 220\,e}$$

The measurement was done at typical 10kHz count rates and anode currents of few nA. The current *I* is the difference of the current with the active source and the current measured without source (dark current). The determined gas gain *G* is shown on the right panel of Figure 8. The gain increases exponentially with the anode voltage, as expected. The mean amplitude using beta source can be normalized to the overlapping voltage region, allowing one to extend the gain measurement range. One has to note that the chamber was already fairly efficient at 1400V, and stable even up to 1700V – that is, a broad, 300V range is offered to find optimal detector working point.

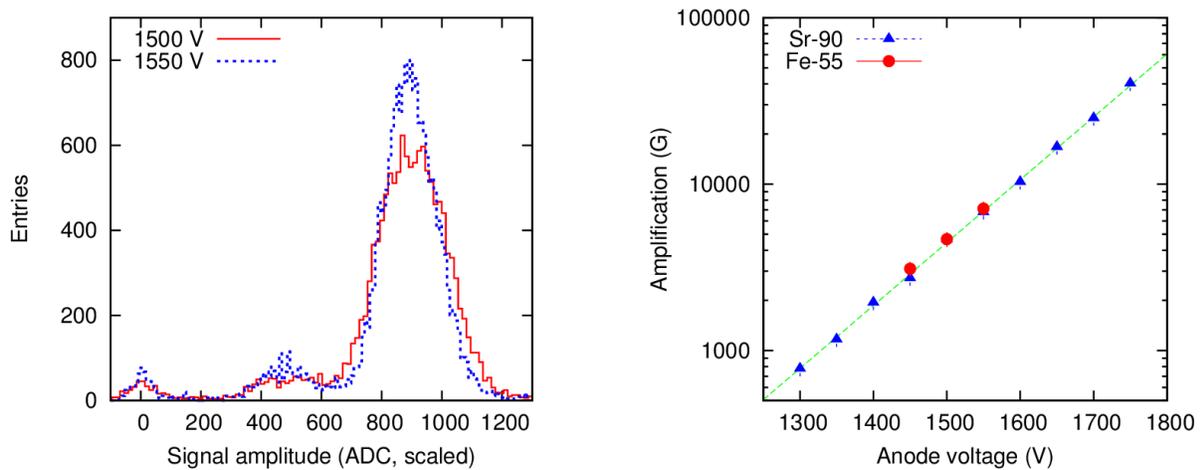

Figure 8. Left: using an $^{55}$Fe source, the signal shape has a well determined peak at the emitted gamma ray energy of 5.9 keV. Data at different voltages are scaled horizontally to demonstrate shape similarity. Right: the amplification gain G has been determined using $^{55}$Fe, and shows consistent gain evolution also with $^{90}$Sr beta signals. The continuous line is an exponential function, drawn to guide the eye.

## 4. Detector performance studies for cosmic muon detection

Cosmic particles offer a possibility to test detectors without radioactive sources. For cosmic muon detection, avalanche amplification gains of the order of $10^4$ are sufficient (1600-1700V on the anode wires for the presented geometry). In this case, the signal by traversing muons is very clearly separated from the noise, as shown on the left panel of Figure 9. For such a measurement, the trigger (start) signal was extracted from the sense wires such that a coincident set of pulses were detected: if within a time window of 2µs (corresponding to the pulse width observed in Figure 6) all of the signals from the four detectors were above a predefined threshold, then it was defined as a "muon event". The scheme is illustrated on the right panel of Figure 9. For muon events the signal peak amplitude was measured, and is shown on the left panel of Figure 9 for each of the individual detectors.

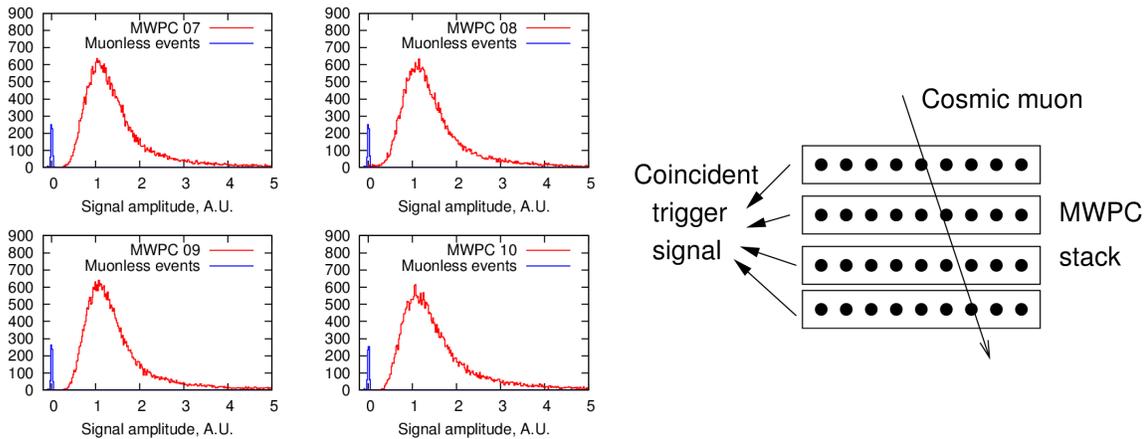

Figure 9. Left: Signal amplitudes in a 4-chamber setup, with or without a traversing muon. Right: measurement scheme for cosmic muon detection, with trigger (start) signal determined by simultaneous signals observed in a number of detectors.

One can conclude that there is a very clear separation between the signal (broad and asymmetric signal, often called "Landau distribution" in case of high energy particle detectors) and noise. Such detectors can find applications as cosmic muon telescopes in various underground applications [7,8], considering the fact that a rather broad range of amplification gain is usable (from few $10^3$ up to $10^5$) for fully efficient detection. The detector is sensitive to any ionizing particles which crosses it: the minimum energy necessary is in the order of 30MeV for electrons or muons. If for specific applications it is useful to increase this threshold, such as the case of muons (with mean energy of order of a GeV), then absorber layers may be added to the detector system.

## 5. Visualization with an LED-array

With a trigger defined by the coincidence of anode wire signals, the pulses induced on the field shaping wires can be picked up by connecting a sensitive amplifier to each of these wires. This principle has been used for building tracking chambers with two dimensional position sensitivity [6]. In fact, if the signal is not read out by a computer for later analysis, but latched and transferred to an LED, a spectacular visualization of the muon trajectories becomes possible. Figure 10 shows the circuit diagram for such a single channel of an amplifier-latch card, which needs to receive a trigger pulse to latch the discriminated input signal and forward it to the LED. Note that the the input connections from individual field shaping wires are shown also in Figure 4.

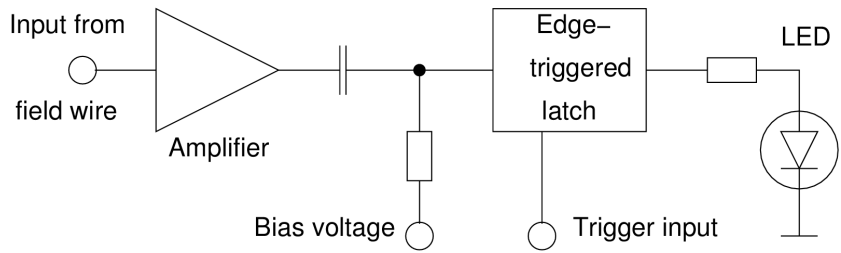

Figure 10. Electric circuit diagram for one channel connected to a field shaping wire. Here an edge-triggered shift register may be used as combined discriminator and latch.

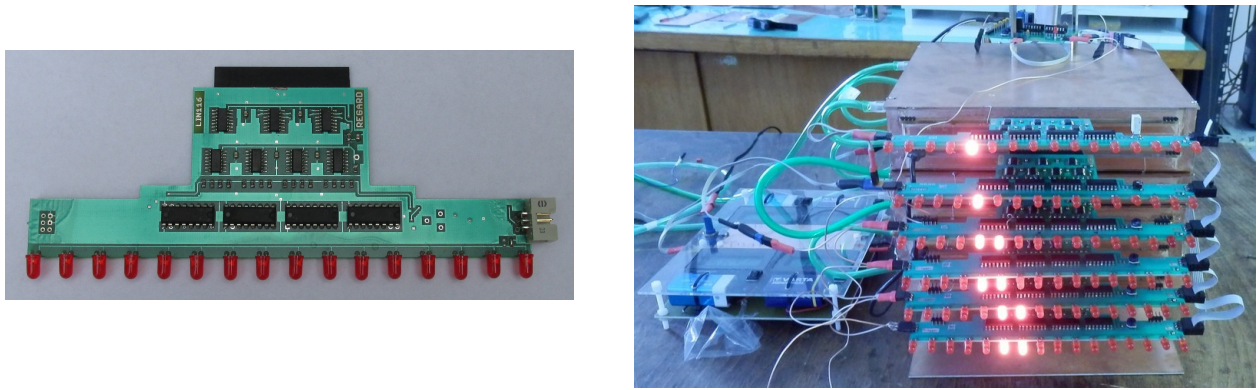

Figure 11. Left: LED display card for signal visualization, with one channel corresponding to one input from a field shaping wire. Right: Clear trajectories are apparent, initiated by traversing cosmic muons.

Once a coincident trigger signal is received, the signal pulses from the field shaping wires display the straight particle trajectory, well visible by human observers. Given the typical ground level muon fluxes, in this 20cm detectors about 1-2 clear cosmic events appear per second.

## 6. Conclusions

The design and construction of classical multi-wire proportional chambers have been presented, which show excellent detection efficiency, moderate position sensitivity and highly stable operation through a broad amplification range. The smaller version of the detectors, in 20cm by 20cm size, may be constructed within classroom conditions involving motivated undergraduate or graduate students, and equipped with an LED display it makes an excellent demonstration device for the existence of highly penetrating cosmic muons. In larger versions a similar MWPC design may be part of cosmic muon telescope systems featuring a cost efficient tracking solution.

**Acknowledgements**

This work was supported by the "Momentum" Programme of the Hungarian Academy of Sciences (LP2013-60) and the "Útravaló – Út a tudományhoz" Programme of the Hungarian Ministry of Human Resources. We wish to acknowledge the contribution of students from the Mechatronika Secondary School in Budapest, as well as members of the REGARD group at the Wigner Research Centre for Physics (Budapest).